\documentstyle[12pt]{article}
\newcommand\be{\begin{equation}}
\newcommand\ee{\end{equation}}
\newcommand\bea{\begin{eqnarray}}
\newcommand\eea{\end{eqnarray}}
\newcommand\ket[1]{|#1\rangle}

\newcommand\braket[2]{\langle #1|#2\rangle}
\newcommand{\fatalpha}{{\bf \alpha \kern -0.44em \alpha}}
\newcommand{\fatsigma}{{\bf \sigma \kern -0.54em \sigma}}
\newcommand{\tpchi}{{\bf \chi \kern -0.35em \chi}}
\newcommand{\llambda}{{\bf \lambda \kern -0.45em \lambda}}

\renewcommand{\theequation}{\arabic{equation}}
\renewcommand{\theequation}{\thesection-\arabic{equation}}
\bibliography{plain}
\title{\bf Continuous-time quantum walks on
semi-regular spidernet graphs via quantum probability
theory}\vspace{20mm}
\author{ S. Salimi
  \thanks{Corresponding author:  E-mail addresses: shsalimi@uok.ac.ir}
 \\ {\small Department of Physics,
University of Kurdistan, P.O.Box 66177-15175 , Sanandaj, Iran.}}
\pagebreak

\vspace{20mm}
\begin{document}
\maketitle \vspace{15mm}
\begin{abstract}
{\bf{We analyze continuous-time quantum and classical random walk on
spidernet lattices. In the framework of Stieltjes transform, we
obtain density of states, which is an efficiency measure for  the
performance of classical and quantum mechanical transport processes
on graphs, and calculate the spacetime  transition probabilities
between two vertices of the lattice. Then we analytically show that
there are two power law decays $\sim t^{-3}$ and $\sim t^{-1.5}$ at
the beginning of the transport for  transition probability in the
continuous-time quantum and classical random walk respectively. This
results illustrate the decay of quantum mechanical transport
processes is quicker than  that of the classical one. Due to the
result, the characteristic time $t_c$, which is  the time when the
first maximum of the probabilities occur on an infinite graph,  for
the  quantum walk is shorter than that of the classical walk.
Therefore, we can interpret that the quantum transport speed on
spidernet is faster than that of the classical one. In the end, we
investigate the results by numerical analysis for two examples.}}


{\bf PACs Index: 03.65.Ud }
\end{abstract}

\vspace{70mm}
\newpage
\section{Introduction}
Quantum walks were introduced in the early 1990s by Aharonovich,
Davidovich and Zaggury \cite{adz}. Since then the topic has
attracted considerable  interest. The continuing attraction can be
traced back to at least two reasons. First, the quantum walk is of
sufficient interest in its own right because there are fundamental
differences compared to the classical random walk. Next, quantum
walks offer quite a number of possible applications. One of the best
known is the link between quantum walks and quantum search
algorithms which are superior to their classical
counterparts\cite{ccdfgs, Kempe}. Similar to classical random walk
there are two types of quantum walks, discrete and continuous time
\cite{adz,Kempe1}. A study of quantum walks on a simple graph is
well known in physics (for more details see \cite{fls}). Recent
studies of quantum walks on more general graphs were described in
\cite{ccdfgs, fg,cfg,abnvw,aakv,mr,kem,khovi, konno2}. Some of these
works study the problem in the important context of algorithmic
problems on graphs and suggest that quantum walk is a promising
algorithmic technique for designing future quantum algorithms. One
approach for investigation of continuous-time quantum walk (CTQW) on
graphs is using the spectral distribution associated with the
adjacency matrix of graphs \cite{js,jsa, jsas, konno1,
salimi1,salimi2}. Authors in Refs.\cite{js,jsa} have introduced a
new method for calculating the probability amplitudes of quantum
walk based on spectral distribution. In this method a canonical
relation between the Fock space  of stratification graph and set of
orthogonal polynomials has been established which leads to obtain
the probability measure (spectral distribution) of adjacency matrix
graph. The method of spectral distribution only requires simple
structural data of graph and allows us to avoid a heavy
combinational argument often necessary to obtain full description of
spectrum of the adjacency matrix.

{\bf{In fact the dendrimers play an important role in  recent
researches, both experimentally and theoretically (e.g. see
Ref.\cite{vogtle}). Dendrimers are hyperbranched macromolecules with
very regular structure, and are important in  drug delivery.
Therefore, that part of theoretical researches depending on probe
transport process is interesting. Since dendrimers are synthesized
in a self-similar fashion by hierarchically growing  dendrimer from
a core, they can be described by spidernet graphs. In  recent years,
the Mulken, Bierbaum and Blumen studied the coherent transport on
dendrimers by CTQW \cite{OMV}. Dendrimers may have the spidernet
lattices structure. As an  example, we can mention the dendrimers
that the authors of Ref.\cite{OMV} considered are spidernet lattices
with parameters $a=b=3; c=2$ , i.e. $S(3,3,2)$}}.  Therefore, in
this paper we study CTQW and continuous-time random walk (CTRW) on
spidernet lattices. For this work, by using the Stieltjes transform,
we obtain spectral distribution (density of states) that  is an
efficiency measure for the performance of the classical and quantum
mechanical transport processes on graphs \cite{mulken}.
{\bf{Considering the results of Ref. \cite{OMV, mulken, mulken1},
one way of quantifying the global efficiency of classical and
 quantum walk is  the average probability of a walker to return
to or stay at the starting point. In the classical case this
quantity is equal to $p_0(t)$ (transition probability at the
starting point at the time $t$ ) and in the quantum mechanical case
 $|q_0(t)|^2$ is the lower bound for the average probability. Then
we  analytically obtain  these quantities for CTQW and CTRW  on
spidernet graphs as $\sim t^{-3}$ and $\sim t^{-3/2}$  ,
respectively. Since the decay of the quantum mechanical average
probability  is quicker than that of the  classical one, the quantum
walk on the spidernet graphs is more efficient than the classical
random walk. In the end, by numerical analysis we confirm these
results for CTQW and CTRW  on two spidernet graphs $S(4,6,3)$ and
$S(a,a,a-1)$. As one can see from the figures,  there are power law
decay probabilities  ($\sim t^{-1.5}$) and $\sim t^{-3}$ at the
beginning of the transport for transition probability for CTRW and
CTQW, respectively. Due to results of this paper, since the
characteristic time $t_c$ depends on the decay of the average
probability ( $t_c$ is the time when the first maximum of the
probabilities occurs on an infinite graph, such that this definition
is held both for the classical and quantum transport. For the
classical transport, there is only one maximal value and the
characteristic time corresponds to the time when the equipartitioned
probability $1/N$ is reached  on finite graphs \cite{xu}. ) the
$t_c$ for the quantum walk is shorter than the classical
counterpart. Therefore, one can interpret that the quantum transport
speed on spidernet is faster than that of the classical one. }}

The organization of the paper is as follows: we give a brief review
of stratification, quantum decomposition and spidernet graph in
Section $2$. Section $3$ is devoted to study CTQW and CTRW on
spidernet graphs via quantum probability theory and try to
investigate CTQW and CTRW on their graphs. In the conclusion we
summarize the obtained results and discuss possible development.
Finally, in the appendix the determination of spectral distribution
associated with adjacency matrix  by Stieltjes transform is derived.

\section{Stratification, Quantum decomposition and Spidernet graph}
Let $V$ be a non-empty set and $E$ be a subset of $\{\{\alpha,
\beta\}| \alpha, \beta\in V \ \mbox{for}\ \alpha\neq \beta \}$. The
pair $G=(V,E)$ is called a graph, where elements of $V$ and $E$ are
 vertices and edges of graph, respectively. We say that two
vertices of $\alpha$ and $\beta$ are adjacent if $\{\alpha,
\beta\}\in E$ and write $\alpha\sim \beta$. A finite sequence
$\alpha_0, \alpha_1,..., \alpha_n$ is said a walk of length $n$ if
$\alpha_k\sim \alpha_{k+1}$ for $k=0, 1, ..., n-1$. A graph is
called connected if any pair of distinct vertices is connected by a
walk. The degree or valency of a vertex $\alpha\in V$ is defined by
\begin{equation}
\kappa(\alpha)=|\{\beta \in V|\ \beta\sim \alpha \}|
\end{equation}
where $|.|$ denote the cardinality. For a graph $G$  the adjacency
matrix $A$ is given by
\[
A_{\alpha \beta} = \left\{
\begin{array}{ll}
1 & \mbox{if $ \alpha\sim \beta$}\\
0 & \mbox{otherwise.}
\end{array}
\right.
\]

Obviously, (i) $A$ is a symmetric (ii)  elements of $A$  take a
value in $\{0, 1\}$ (iii) diagonal elements of $A$ are $0$.
Conversely, for a non-empty set $V$, a structure graph is uniquely
determined by a such matrix which indexed by $V$. On the other hand,
$A$ is considered as an operator acting on the Hilbert space
$l^2(V)$ in such a way that
$$
A\ket{\alpha}=\sum_{\alpha\sim \beta}\ket{\beta}, \;\;\;\;\alpha\in
V,
$$
where $\{\ket{\alpha}|\ \alpha \in V\}$ forms a complete orthogonal
basis of $l^2(V)$.

 The
stratification is introduced \cite{js,nob, hnob} by taking $o$ as
the origin (initial site) and have
\begin{equation}\label{v1}
V=\bigcup_{k=0}^{\infty}V_k,\;\;\;\;\;\; V_i=\{\alpha\in V|\
\partial(o,\alpha)=k\},
\end{equation}
where $\partial(\alpha, \beta)$ stands for the length of the
shortest walk connecting $\alpha$ and $\beta$. According to the
stratification (\ref{v1}), we define a unit vector by
\begin{equation}
\ket{\phi_{k}}=\frac{1}{\sqrt{|V_k|}}\sum_{\alpha\in V_{k}}\ket{k,
\alpha},
\end{equation}
where $\ket{k, \alpha}$ denotes the eigenket of the $\alpha$-th
vertex at the stratum $k$ and let $\Gamma(G)$ the closed subspace of
$l^2(V)$ be spanned by $\{\ket{\phi_{k}}\}$. {\bf{Moreover, the
stratification (\ref{v1}) give rise to define three matrices
$A^\varepsilon, \ \varepsilon\in \{+,-,0\}$, as follows \cite{hnob}:
for $\alpha\in V_k, \ k=0,1,2,...  $
\[
(A^\varepsilon)_{\beta \alpha}=\left\{\begin{array}{ll}
A_{\beta \alpha} & \mbox{if $ \beta \in V_{k+\varepsilon}$}\\
0 & \mbox{ otherwise.}
\end{array}
\right.
\]
where $k+\varepsilon=n+1, n-1, n$ according as $ \varepsilon\in
+,-,0$. Therefore, the adjacency matrix $A$ is decomposed as
\begin{equation}\label{qd1}
A=A^{+}+A^{-}+A^0.
\end{equation}
This is called quantum decomposition of $A$ associated with the
stratification (\ref{v1}). }} Now,  according to the stratification
(\ref{v1}), for $\alpha \in V_k$ we set
\begin{equation}
\omega_{\varepsilon}(\alpha)=|\{\beta\in V_{k+\varepsilon},
\alpha\sim \beta\}|, \qquad \varepsilon\in \{+,-,0\}
\end{equation}
where $k+\varepsilon=n+1, n-1, n$ according as $ \varepsilon\in
+,-,0$. The degree or valency of $\alpha \in V$ is

\[
\kappa(\alpha)=\omega_{+}(\alpha)+\omega_{0}(\alpha)+\omega_{-}(\alpha)
=\left\{\begin{array}{ll}
a & \mbox{ $ \alpha=o$}\\
b & \mbox{$ \alpha\neq o$.}
\end{array}
\right.
\]

We consider the integers $a, b, c$ such that $a\geq 1,\ b\geq 2 $
and $1\leq c\leq b-1$.  A spidernet is a graph which satisfies the
following conditions:
$$
  \omega_{+}(o)=a \quad
  \omega_{-}(o)=0 \quad
  \omega_{0}(o)=0
  $$
\begin{equation}\label{spidernet}
  \omega_{+}(\alpha)=c\quad
  \omega_{-}(\alpha)=1\quad
  \omega_{0}(\alpha)=b-1-c
 \quad \mbox{for}\ \alpha\neq o,
\end{equation}
where it is denoted by $S(a,b,c)$, for example see Fig.1 (for more
details see Refs.\cite{hnob}). Spidernet is not necessarily a
regular graph so it is called a semi-regular graph (for more details
see Refs.\cite{hnob,urakawa}). It is easy to show, by using
(\ref{spidernet}), the number of vertices in  strata as
\begin{equation}\label{starta}
|V_0|=1, \quad |V_k|=ac^{k-1}, \quad k=1,2,3,....
\end{equation}
Then by using Eqs.(\ref{spidernet}) and (\ref{starta}), one can
obtain
\begin{equation}\label{v5}
A^{+}\ket{\phi_{0}}=\sqrt{a}\ket{\phi_{1}},\quad
A^{+}\ket{\phi_{k}}=\sqrt{c}\ket{\phi_{k+1}},\quad k\geq 1
\end{equation}
\begin{equation}\label{v6}
A^{-}\ket{\phi_{0}}=0, \quad
A^{-}\ket{\phi_{1}}=\sqrt{a}\ket{\phi_{0}},\quad
A^{-}\ket{\phi_{k}}=\sqrt{c}\ket{\phi_{k-1}}, \;\;\;\ k\geq 2
\end{equation}
\begin{equation}\label{v7}
A^{0}\ket{\phi_{k}}=(b-1-c)\ket{\phi_{k}}, \quad k\geq 1.
\end{equation}
Two Szeg\"{o}- Jacobi sequences $\{\omega_k\}_{k=1}^{\infty}$ and
$\{\alpha_k\}_{k=1}^{\infty}$  can obtain to take into consideration
above equation as following \cite{hnob}
\begin{equation} \omega_1=a, \ \omega_2=\omega_3=\cdots=c,\quad
\alpha_1=0, \quad \alpha_{k+1}=b-1-c \quad k\geq 1,
\end{equation}
where  $(\Gamma(G), A^+, A^-, A^o )$ is an interacting Fock space
associated with Szeg\"{o}- Jacobi sequences $\{\omega_k,
\alpha_k\}$.
\section{CTQW and CTRW on spidernets via quantum probability theory }
We start our discussion by considering a walk on general graphs.
Classically, the evolution of CTRW is governed by Kolmogorov's
equation (master equation)\cite{Weiss, Kampen},
\begin{equation}
\frac{dP_{i,j}}{dt}=\sum_{k} H_{ik}P_{k,j}
\end{equation}
where $H$ is Hamiltonian of the walk and $P_{i,j}$ is the
conditional probability to find the CTRW on vertex $i$ at time $t$
when the walk starting in the vertex $j$. It is natural to choose
the Laplacian of the graph, defined as $L=D-A$, as  Hamiltonian of
walk where $D$ is a diagonal matrix with entries as
$D_{jj}=deg(\alpha_j)$. Then the solution of the above equation is
\begin{equation}
P_{k,j}=\langle k|e^{tH}|j\rangle.
\end{equation}
Quantum mechanically, the evolution of CTQW is obtained by replacing
Kolmogorov's equation with Schr\"{o}dinger's equation
\begin{equation}
i\hbar\frac{d\ket{\phi(t)}}{dt}= H\ket{\phi(t)}.
\end{equation}
where we assume $\hbar=1$ and $\ket{\phi_0}$ is the initial
amplitude wave function of the particle. The solution of the above
equation is given by
\begin{equation}
\ket{\phi(t)}=e^{-itH}\ket{\phi_0}
\end{equation}
On $s$-regular graphs $D=sI$, then we get
\begin{equation}
e^{-itH}=e^{-it(A-sI)}=e^{itsI}e^{-itA}.
\end{equation}
This introduces an irrelevant phase factor in the wave evolution,
hence we can consider $H=A$.

{\bf{In the study of CTQW on graphs, the spectral distribution or
density of states of the system, $H$ contains  essential information
about the system \cite{js} (in fact, the spectral distribution is an
efficiency measure for the performance of classical and quantum
mechanical transport processes  on graphs \cite{mulken} ) and by
definition the spectral distribution is  a probability distribution
$\mu$ uniquely specified by}}
\begin{equation}
\langle H^m\rangle= \langle \phi_0| H^m|\phi_0\rangle=\int x^m
\mu(dx), \quad m=0,1,2,...,
\end{equation}
where, according to \cite{js, jsa,nob}, $\langle H^m\rangle$
coincides with the number of $m$-step walks starting and terminating
at the origin site $o$. For analyzing the spectral distribution
$\mu$ of the adjacency matrix $A$, we use the method of quantum
decomposition method which is a powerful tool. The spectral
distribution $\mu$ is determined by applying the canonical
isomorphism from the interacting Fock space onto the closed linear
span of orthogonal polynomials determined by  Szeg\"{o}- Jacobi
sequences $\{\omega_k, \alpha_k\}$. In fact the determination of
$\mu$ is the main problem in the spectral theory of operators, where
in the case is quite possible by using the Stieltjes method, as it
is explained in appendix A. Then by using the quantum decomposition
relations (\ref{qd1})-(\ref{v7}) and the recursion relation of
polynomials $P_n(n)$(\ref{op}), the other matrix elements as
\begin{equation}\label{prob1}
\langle \phi_k|
H^m|\phi_0\rangle=\frac{1}{\sqrt{\omega_1\omega_2...\omega_k}}\int
x^m P_k(x)\mu(dx), \quad m=0,1,2,....
\end{equation}
Then by using Eq.(\ref{prob1}), the classical transition probability
and quantum mechanical transition amplitude  to go from the initial
site $o$ at time $0$ to the stratum $\ket{\phi_k}$ in time $t$ are
given by
$$
p_0(t)=\langle \phi_0| e^{tH}|\phi_0\rangle=\int e^{tx} \mu(dx),
$$
\begin{equation}\label{classic1}
p_k(t)=\langle \phi_k|
e^{tH}|\phi_0\rangle=\frac{1}{\sqrt{\omega_1\omega_2...\omega_k}}\int
e^{tx} P_k(x)\mu(dx) \quad \mbox{for}\ k\geq 1,
\end{equation}
and
$$
q_0(t)=\langle \phi_0| e^{-itH}|\phi_0\rangle=\int e^{-itx}\mu(dx)
$$
\begin{equation}\label{prob2}
q_k(t)=\langle \phi_k|
e^{-itH}|\phi_0\rangle=\frac{1}{\sqrt{\omega_1\omega_2...\omega_k}}\int
e^{-itx} P_k(x)\mu(dx) \quad \mbox{for}\ k\geq 1,
\end{equation}
respectively, where $|q_{k}(t)|^2$ is the  transition probability of
the quantum walk at the stratum $k$ at time $t$. The conservation of
probability $\sum_{k}|q_{k}(t)|^2=1$ ($\sum_{k}p_{k}(t)=1$) follows
immediately from Eq. (\ref{prob2}) by using the completeness
relation of orthogonal polynomials $P_n(x)$. In the appendix $A$
reference \cite{js} is provided the walker has the same transition
probability  at the all sites belonging to the same stratum, i.e.,
we have $|q_{ik}(t)|^2=\frac{|q_k(t)|^2}{|V_k|}, \ \mbox{for}\ i\in
V_k$ ($p_{ik}(t)=\frac{p_k(t)}{|V_k|}, \ \mbox{for}\ i\in V_k$),
where $|q_{ik}(t)|^2$ ($p_{ik}(t)$) denotes the transition
probability of the quantum walk (classical walk) at the $i$-th
vertex of $k$-th stratum $V_k$. Investigation of CTQW via spectral
distribution method, which is introduced as a new method for
calculating the probability amplitudes quantum walk (for more
details see \cite{js} ), allows us to avoid a heavy combinational
argument often necessary to obtain full description of spectrum of
the Hamiltonian.

We can now  investigate CTQW on spidernets. For calculating CTQW we
need the Stieltjes transform $G_{\mu}(z)$ which takes the following
form
\begin{equation}\label{q11}
G_{\mu}(z)=
\frac{1}{z-\frac{a}{z-(b-1-c)-\frac{c}{z-(b-1-c)-\frac{c}{z-(b-1-c)-\frac{c}{\ddots}}}}}
\end{equation}
In order to evaluate the continued fraction, we need  firstly to
evaluate the following infinite continued fraction defined as
\begin{equation}
\tilde{G}_{\mu}(z)=
\frac{c}{z-(b-1-c)-\frac{c}{z-(b-1-c)-\frac{c}{z-(b-1-c)-\frac{c}{z-(b-1-c)-\frac{c}{\ddots}}}}}=\frac{c}{z-(b-1-c)-\tilde{G}_{\mu}(z)},
\end{equation}
where by solving the above equation, we have
\begin{equation}\label{q12}
\tilde{G}_{\mu}(z)=\frac{1}{2}(z-(b-1-c)-\sqrt{(z-(b-1-c))^2-4c}).
\end{equation}
By substituting (\ref{q12}) into (\ref{q11}), we obtain the
following expression for the Stieltjes transform of $\mu$
\begin{equation}\label{q1}
G_{\mu}(z)=
\frac{1}{z-\frac{a}{z-(b-1-c)-\tilde{G}_{\mu}(z)}}=\frac{1}{2}\frac{a(b-1-c)+(2c-a)z-a\sqrt{(z-(b-1-c))^2-4c}}{a^2+((c-a)z+a(b-1-c))z}.
\end{equation}
Finally, by applying Stieltjes inversion formula, we acquire the
absolutely continuous part of spectral distribution $\mu$ as follows
\begin{equation}\label{q2}
\mu(x)=\frac{1}{2\pi}\frac{a\sqrt{4c-(x-(b-1-c))^2}}{a^2+((c-a)x+a(b-1-c))x}
\ ; \qquad -2\sqrt{c}+(b-1-c)\leq x\leq 2\sqrt{c}+(b-1-c).
\end{equation}
{\bf{Referring to the results of  Refs. \cite{OMV, mulken, mulken1},
 $|q_0(t)|^2$ (return probability at the initial vertex) is the
lower bound of the average probability to be still or again at the
initially excited vertex for quantum walker and $p_0(t)$ is the
average probability to return the initial vertex for classical
walker. Since the lower bound in the CTQW oscillates, due to unitary
time evolution, one can use the envelope of this oscillations as a
measure for quantifying the efficiency (for more detail, see Ref.
\cite{mulken1}).   Now, by considering Refs. \cite{OMV, mulken,
mulken1}, we are in the position to quantify the efficiency of
classical and quantum mechanical transport processes on these
graphs. For doing this work, we study the asymptotic expansion of
integral $p_o(t)$ in the Eq.(\ref{classic1}) for small $x$ and large
$t$. The spectral distribution  of Eq.(\ref{q1}) for small $x$ is
$\mu(x)\sim x^{-1/2}(d+x)^{-1/2}\sim x^{-1/2}(1-\frac{1}{2}x)$, then
we have
\begin{equation}
p_0(t)\sim t^{-3/2}.
\end{equation}
To obtain this result we use the Ref.\cite{bender}(for more details
see chapter $6$). Therefore, one can obtain  the lower bound of
quantum mechanical for transport processes as
\begin{equation}
|q_0(t)|^2\sim t^{-3},
\end{equation}
(for more details see Ref. \cite{mulken}). Since the decay of
$|q_0(t)|^2$ is much quicker than that of $p_0(t)$, the quantum walk
on this graph  is more efficient than the classical random walk
\cite{OMV, mulken1}. Now, by numerical analysis,  we confirm this
results for CTQW and CTRW  on two infinite graphs which can be
obtained from spidernet graph by an appropriate choice of $a, b,
c$.}}

{\textbf{Example A.}} \\
In this example we consider spidernet in Fig.1, i.e.,  $S(4,6,3)$.
Since this graph is not regular it is sufficient to place
$\acute{\alpha_k}=D_{kk}-\alpha_k $ (i.e., $\acute{\alpha_1}=a=4,
\acute{\alpha_2}=\acute{\alpha_3}=\cdots=c+1=4 $) in the
relation(\ref{q1}). Then the spectral distribution takes the
following form
\begin{equation}
\mu(x)=\frac{1}{2\pi}\frac{4\sqrt{12-(x+4)^2}}{-8x-x^2}, \qquad
-4-2\sqrt{3}\leq x\leq -4+2\sqrt{3}.
\end{equation}
The transition probability and amplitudes of the walker at the
stratum  $k$ at the time $t$, for classical and quantum, are given
by
$$
p_0(t)=-\frac{2}{\pi}\int_{-4-2\sqrt{3}}^{-4 +2\sqrt{3}} e^{tx}
\frac{\sqrt{12-(x+4)^2}}{8x+x^2}dx,
$$$$
p_k(t)=-\frac{1}{\pi}\frac{1}{\sqrt{3^{k-1}}}\int_{-4-2\sqrt{3}}^{-4+2\sqrt{3}}
e^{tx} P_k(x)\frac{\sqrt{12-(x+4)^2}}{8x+x^2}dx, \quad \mbox{for} \
k\geq 1,
$$
$$
q_0(t)=-\frac{2}{\pi}\int_{-4-2\sqrt{3}}^{-4 +2\sqrt{3}} e^{-itx}
\frac{\sqrt{12-(x+4)^2}}{8x+x^2}dx,
$$
\begin{equation}\label{s463}
q_k(t)=-\frac{1}{\pi}\frac{1}{\sqrt{3^{k-1}}}\int_{-4-2\sqrt{3}}^{-4+2\sqrt{3}}
e^{-itx} P_k(x)\frac{\sqrt{12-(x+4)^2}}{8x+x^2}dx, \quad \mbox{for}
\ k\geq 1,
\end{equation}
respectively. {\bf{Then, for analyzing the time dependent of the
transition probability,  we calculate numerically $p_0(t),
|q_0(t)|^2, p_1(t)$ and $|q_1(t)|^2$ of Eq.(\ref{s463}). Figs.2 and
3 show the return probability for CTRW and CTQW on the initial site
$o$ of the graph. For CTRW, there is a power law decay $\sim
t^{-1.5}$ at the beginning of the transport, but after some time,
$p_0$ reaches a constant value $\frac{1}{N}$ ( $N$ is  total number
of the vertices of the graph i.e, $N=|V|$). Since the size of the
graph is infinite this constant goes to the zero, as  shown in
Fig.2. }}

Also in  Fig.3, the dashed curve indicates the quantum mechanical
return probability on the initial site $o$ of the graph. The dashed
line shows the scaling behavior as $\sim t^{-3}$, but at large
times, the return probability oscillates frequently and approaches
to zero. This property indicates that the walk escapes from the
starting site $o$.

In order to compare the transport speed on this graph, we define the
characteristic time $t_c$ as the time when the  first maximum of the
probabilities occurs on an infinite graph. Such definition is held
for both the classical and quantum mechanical transport. For the
classical transport, there is one maximal value and the
characteristic time corresponds to the time when the  probability is
reached uniform probability on finite graph. Figure $3$ indicates
that the characteristic time $t_c$ for the quantum walk is shorter
than that of the  classical one. Therefore, we can interpret that
the quantum transport speed on spidernet $S(4,6,3)$ is faster than
that of the classical one. The different behavior of the transport
speeds between the quantum and classical random walk is striking
characteristic that distinguishes the two transport processes.

 {\textbf{Example B.}}\\
 First we consider $b=a$ and $c=a-1$. With this choice the spidernet
$S(a,a,a-1)$ is  graph with degree $\kappa=a$  and $\mu$ obtains
from Eq.(\ref{q1}) as
\begin{equation}
\mu(x)=\frac{1}{2\pi}\frac{a\sqrt{4(a-1)-x^2}}{a^2-x^2}, \qquad
-2\sqrt{a-1}\leq x\leq 2\sqrt{a-1},
\end{equation}
where this type of measure was first obtained by Kesten
\cite{kesten} in classical random walk with a different method. By
using the Eq.(\ref{prob2}), the transition probability and
amplitudes for observing walker at the stratum $k$ at the time $t$,
for classical and quantum, are given by
$$
p_0(t)=\frac{a e^{-at}}{2\pi}\int_{-2\sqrt{a-1}}^{2\sqrt{a-1}}
e^{tx} \frac{\sqrt{4(a-1)-x^2}}{a^2-x^2}dx.
$$
$$
p_k(t)=\frac{ae^{-at}}{2\pi}\frac{1}{\sqrt{a(a-1)^{k-1}}}\int_{-2\sqrt{a-1}}^{2\sqrt{a-1}}
e^{tx} P_k(x)\frac{\sqrt{4(a-1)-x^2}}{a^2-x^2}dx, \quad \mbox{for} \
k\geq 1,
$$
$$
q_0(t)=\frac{a}{2\pi}\int_{-2\sqrt{a-1}}^{2\sqrt{a-1}} e^{-itx}
\frac{\sqrt{4(a-1)-x^2}}{a^2-x^2}dx.
$$
\begin{equation}\label{tree}
q_k(t)=\frac{a}{2\pi}\frac{1}{\sqrt{a(a-1)^{k-1}}}\int_{-2\sqrt{a-1}}^{2\sqrt{a-1}}
e^{-itx} P_k(x)\frac{\sqrt{4(a-1)-x^2}}{a^2-x^2}dx, \quad
\mbox{for} \ k\geq 1,
\end{equation}
respectively. particularly, when $a=2$ the network $S(2,2,1)$
corresponds to a line graph which the transition amplitudes
 can be rewritten as $q_{0}(t)=J_0(2t)$
($p_{0}(t)=e^{-2t}I_0(2t)$) and $q_{k}(t)=\sqrt{2}i^kJ_k(2t)$
($p_{k}(t)=e^{-2t}\sqrt{2}I_k(2t)$) for $k\geq 2$ where $J_k$ and
$I_k$ are the Bessel and modified Bessel function of the first kind,
respectively. This is consistent with the result in Ref.\cite{js}.
For analyzing the time dependent of probability the observing walk
we calculate numerically $|q_k(t)|^2, p_k(t)$ of  Eq.(\ref{tree}).
The return probability at the initial vertex is a good measure to
quantify the efficiency of the transport\cite{mulken}.

{\bf{Having obtained transition probability for arbitrary  spidernet
$S(a,a,a-1)$ we investigate numerically CTRW and CTQW on its graph
for $a=3,4$.   Figs. $4,5$ show the transition probability for CTRW
on spidernet with $a=3,4$, respectively. For CTRW, there is a power
low decay ($\sim t^{-1.5}$) at the beginning of the transport, but
after some time the transition probability reach a constant value.
Also, the figures $6, 7$ show the transition probability for CTQW
this graph with with $a=3,4$, respectively. In this case the dashed
line shows the scaling behavior $\sim t^{-3}$, and the return
probability oscillate frequently and approach to zero which this
property indicate the walk escape form a starting site $o$. This
figures indicate that the characteristic time $t_c$ for quantum walk
is shorter than it classical. Therefore, we can interpret which the
quantum transport speed on spidernet is faster than that of the
classical one.}}

Finally, for large $a$ (i.e., $a\longrightarrow\infty$) we discuss
this question as a quantum central limit theorem \cite{salimi2}.
Since $q_k(t)=0$ in the limit $a\longrightarrow\infty$, then we have
normalization Hamiltonian and probability amplitudes as
$$
q_{k}(t)=\lim_{a\longrightarrow\infty}\braket{\phi_k}{e^{-iAt/\sqrt{a}}|\phi_0}=\lim_{a\longrightarrow\infty}
\frac{1}{\sqrt{a(a-1)^k}}\int_{-2\sqrt{a-1}}^{2\sqrt{a-1}}
e^{-itx/\sqrt{a}}P_k(x)\frac{a}{2\pi}\frac{\sqrt{4(a-1)-x^2}}{a^2-x^2}dx
$$
$$
=\lim_{a\longrightarrow\infty}
\frac{1}{2\pi\sqrt{a(a-1)^k}}\int_{-2\sqrt{(a-1)/a}}^{2\sqrt{(a-1)/a}}
e^{-itx} P_k(\sqrt{a}x)\frac{\sqrt{4(a-1)/a-x^2}}{1-x^2/a}dx
$$
$$
=\frac{1}{2\pi}\int_{-2}^{2} e^{-itx} P_{k,\infty}(x)\sqrt{4-x^2}dx
$$
\begin{equation}\label{tree2}
=\frac{2}{\pi}\int_{-1}^{1} e^{-i2xt}
P_{k,\infty}(2x)\sqrt{1-x^2}dx,
\end{equation}
where the polynomials $P_{k,\infty}(x)$ is defined as
\begin{equation}\label{recure1}
P_{k,\infty}(x)=\lim_{a\longrightarrow\infty}
\frac{1}{\sqrt{a(a-1)^k}}P_{k}(x).
\end{equation}
By comparing this recursion relation with Tchebichef polynomials, we
have
\begin{equation}
P_{k,\infty}(x)=U_{k}(x/2),
\end{equation}
where $U_{k}(x)$ is the Tchebichef polynomials of second kind.
Therefore, the probability amplitudes of Eq.(\ref {tree2}) obtain
\begin{equation}
q_{k}(t)=\frac{2}{\pi}\int_{-1}^{1} e^{-i2xt}
U_{k}(x)\sqrt{1-x^2}dx=i^k(k+1)\frac{J_{k+1}(2t)}{t},
\end{equation}
where in this case the return probability is
$|q_{0}(t)|^2=\left(\frac{J_{1}(2t)}{t}\right)^2$ which the results
are obtained in agreement with Ref.\cite{konno1}. {\bf{In this case,
we see that the spectral distribution obeys  Wigner's semi-circle
law (i.e, $\mu(x)= \frac{1}{\pi}(\sqrt{1-x^2})$) and one can obtain
that the power law decay $\sim t^{-3}$ and $\sim t^{-3/2}$ for the
quantum mechanical transport processes and classical counterpart
(for more details see Ref. \cite{mulken}). }}

\section{Conclusion}
{\bf{In this paper we have studied CTQW and CTRW on spidernet
lattices. For this work, by using the Stieltjes transform, we have
obtained spectral distribution (density of states) that it is a
efficiency measure for the performance of classical and quantum
mechanical transport processes on graphs. The we have analytically
obtained the power laws $\sim t^{-3}$ and $\sim t^{-3/2}$ for CTQW
and CTRW on spidernet graphs, respectively. By numerical analysis
for two examples $S(4,6,3)$ and $S(a,a,a-1)$, we have confirmed
this results. Due to quick decrease of the quantum mechanical
transport processes than that of the  classical one, the quantum
walk on spidernet graphs is more efficient than the classical
counterpart. Also, this results have shown that the characteristic
time $t_c$ (the $t_c$ is the time when the  first maximum of the
probabilities occur on infinite graph ) for quantum walk is
shorter than it classical. Therefore, in this lattices the speed
of quantum transport is faster than that of the classical one.}}

\vspace{1cm} \setcounter{section}{0}
 \setcounter{equation}{0}
 \renewcommand{\theequation}{A-\arabic{equation}}
  {\Large{Appendix A}}\\
\textbf{\large{Determination of spectral distribution by the
Stieltjes transform }}

In this appendix we explain how we can determine spectral
distribution $\mu(x)$ of the graphs, by using the Szeg\"{o}-Jacobi
sequences $(\{\omega_k\},\{\alpha_k\})$. To this aim we may apply
the canonical isomorphism from the interacting Fock space onto the
closed linear span of the orthogonal polynomials determined by the
Szeg\"{o}-Jacobi sequences $(\{\omega_i\},\{\alpha_i\})$. More
precisely, the spectral distribution $\mu$ under question is
characterized by the property of orthogonalizing the polynomials
$\{P_n\}$ defined recurrently by
$$ P_0(x)=1, \;\;\;\;\;\
P_1(x)=x-\alpha_1,$$
\begin{equation}\label{op}
xP_n(x)=P_{n+1}(x)+\alpha_{n+1}P_n(x)+\omega_nP_{n-1}(x),
\end{equation}
for $n\geq 1$.

As it is shown in \cite{tsc}, the spectral distribution ì can be
determined by the following identity:
\begin{equation}\label{v3}
G_{\mu}(z)=\int_{R}\frac{\mu(dx)}{z-x}=\frac{1}{z-\alpha_1-\frac{\omega_1}{z-\alpha_2-\frac{\omega_2}
{z-\alpha_3-\frac{\omega_3}{z-\alpha_4-\cdots}}}}=\frac{Q_{n-1}^{(1)}(z)}{P_{n}(z)}=\sum_{l=1}^{n}
\frac{A_l}{z-x_l},
\end{equation}
where $G_{\mu}(z)$ is called the Stieltjes transform and $A_l$ is
the coefficient in the Gauss quadrature formula corresponding to the
roots $x_l$ of polynomial $P_{n}(x)$ and where the polynomials
$\{Q_{n}^{(1)}\}$ are defined
recurrently as\\
        $Q_{0}^{(1)}(x)=1$,\\
    $Q_{1}^{(1)}(x)=x-\alpha_2$,\\
    $xQ_{n}^{(1)}(x)=Q_{n+1}^{(1)}(x)+\alpha_{n+2}Q_{n}^{(1)}(x)+\omega_{n+1}Q_{n-1}^{(1)}(x)$,\\
    for $n\geq 1$.

Now if $G_{\mu}(z)$ is known, then the spectral distribution $\mu$
can be recovered from $G_{\mu}(z)$ by means of the Stieltjes
inversion formula:
\begin{equation}\label{m1}
\mu(y)-\mu(x)=-\frac{1}{\pi}\lim_{v\longrightarrow
0^+}\int_{x}^{y}Im\{G_{\mu}(u+iv)\}du.
\end{equation}
Substituting the right hand side of (\ref{v3}) in (\ref{m1}), the
spectral distribution can be determined in terms of $x_l,
l=1,2,...$, the roots of the polynomial $P_n(x)$, and  Guass
quadrature constant $A_l, l=1,2,... $ as
\begin{equation}\label{m}
\mu=\sum_l A_l\delta(x-x_l)
\end{equation}
 ( for more details see Refs. \cite{js,jsa,tsc,st}.)

\newpage
{\bf Figure Captions}

{\bf Figure-1:} Spidernet lattice $S(4,6,3)$ {\bf given by Hora et
al. (\cite {hnob}, Fig. 4.3, p. 121)}.

{\bf Figure-2:} The solid and dashed curves  show the return
probability on initial site $o$ and transition probability from
initial site to the first strata for CTRW on spidernet lattice
$S(4,6,3)$, respectively.

{\bf Figure-3:} (a) and (b) show the return probability on initial
site $o$ and transition probability from initial site to the first
strata for CTQW on spidernet lattice $S(4,6,3)$, respectively, in
which the dashed curves indicate the scaling behavior $|q_{0}(t)|^2$
and  $|q_{1}(t)|^2 \sim t^{-3}$.

{\bf Figure-4:} Fig.4 indicates the transition probabilities
$p_{0}(t)$ and $p_{1}(t)$ of CTRQ on $S(a,a, a-1)$, for $a=3$.

{\bf Figure-5:} Fig.5 shows the transition probabilities $p_{0}(t)$
and $p_{1}(t)$ of CTRQ on $S(a,a, a-1)$, for $a=4$.

{\bf Figure-6:}  (a) and (b) indicate the transition probabilities
$|q_{0}(t)|^2$ and $|q_{1}(t)|^2$ for CTQW on $S(a,a, a-1)$, for
$a=3$, in which  the dashed curves show the scaling behavior
$|q_{0}(t)|^2$ and $|q_{1}(t)|^2 \sim t^{-3}$.

{\bf Figure-7:} (a) and (b)  indicate the transition probabilities
$|q_{0}(t)|^2$ and $|q_{1}(t)|^2$  for CTQW on $S(a,a, a-1)$, for
$a=4$, in which  the dashed curves show the scaling behavior
$|q_{0}(t)|^2$ and,$ |q_{1}(t)|^2 \sim t^{-3}$.

\end{document}